# An avalanche-photodiode-based photon-number-resolving detector

*B. E. Kardynał, Z. L. Yuan and A. J. Shields[*]*

Toshiba Research Europe Ltd, 208 Cambridge Science Park, Cambridge CB4 0GZ, United Kingdom

**Avalanche photodiodes are widely used as practical detectors of single photons.[1] Although conventional devices respond to one or more photons, they cannot resolve the number in the incident pulse or short time interval. However, such photon number resolving detectors are urgently needed for applications in quantum computing,[2-4] communications[5] and interferometry,[6] as well as for extending the applicability of quantum detection generally. Here we show that, contrary to current belief,[3,4] avalanche photodiodes are capable of detecting photon number, using a technique to measure very weak avalanches at the early stage of their development. Under such conditions the output signal from the avalanche photodiode is proportional to the number of photons in the incident pulse. As a compact, mass-manufactured device, operating without cryogens and at telecom wavelengths, it offers a practical solution for photon number detection.**

It has been shown that quantum computation can be performed using only linear optical components, single photon sources and detectors that can measure the number of photons in a pulse. The role of photon number resolving detectors in this scheme cannot be overstated since they are used to monitor the operation of the quantum gates. In the original non-linear sign gate proposed by Knill, Laflamme and Milburn,[2] for example, each individual operation can result in 0, 1 or 2 photons in ancillary detectors and successful operation of the gate requires the detection of exactly 0 and 1 photons in these detectors. Another gate of great importance for quantum communication and computation allows extension of the distance over which entanglement can be distributed.[4] This entanglement swapping gate requires post-selection based on exactly one or two-photons registered in a detector. Photon number resolving detectors would also be a very important tool in characterisation of quantum light sources,[7,8] as well as for low light level detection in general.

Avalanche photodiodes (APDs) are most commonly used for single photon detection, as they are simple, robust semiconductor devices with high detection efficiencies and low dark count noise. When the diode is reversed biased above the breakdown voltage, photo-excitation of a single carrier followed by carrier multiplication through repeated impact ionisation results in a macroscopic current that can be easily detected. Detection efficiencies of up to 45% have been reported for telecom-wavelength InGaAs/InP APDs,[9] while values in excess of 65% are common for shorter-wavelength Si devices.[10] Furthermore by integrating the absorption region in a monolithic semiconductor cavity, detection efficiencies exceeding 90% have been demonstrated in linear mode.[11] The detectors can be used even at room temperature although for best performance they are often cooled thermo-electrically to -30°C.

To date it has been thought that APDs are unable to measure the number of photons in a pulse of radiation or a short time interval. Although some alternative technologies have shown limited photon number resolving capability,[12-16] they all require cryogenic cooling and often have low efficiency, a small active area or long integration times. In order to exploit the practical advantages of APDs, some schemes *approximating* photon number resolution were proposed in which an incident pulse is split and registered on multiple detectors or a single detector at different times.[17,18] In the following, we show for the first time that individual APDs can count photons in an incident pulse of laser light with no need for splitting in space or time.

In an InGaAs/InP APD one (or more) photo-excited hole triggers avalanche multiplication in an InP layer subject to a strong electric field. A cascade of impact ionisation events started by the photo-hole(s) results in a current path between the APD electrodes and a self-sustaining avalanche current flows through the device. Once the avalanche is established in the whole multiplication volume of the APD the current saturates at a value which depends on the applied bias and on the total resistance of the circuit, but *does not* depend on the number of photo-generated holes that triggered the process.[19] As the avalanche is self-sustaining the bias must be reduced below breakdown voltage to quench the avalanche.

InGaAs APDs are operated commonly in gated Geiger mode, for which periodic voltage pulses are applied to bias the APD above the breakdown voltage ($V_{BD}$) by an excess bias ($V_{EX}$) for a short duration of several nanoseconds to enable, and then subsequently quench, the avalanche. This mode is well suited to quantum key distribution and other applications using a pulsed source. The photon-induced output signal (measured as voltage drop across a 50 Ω series resistor due to the APD current) is superimposed on the capacitive response of the device to the applied voltage pulses, as shown in Fig.1a. This capacitive signal sets a limit upon the weakest avalanche amplitude that can be detected for conventional operation.

In order to resolve the incident photon number it is essential to measure the current shortly after the onset of avalanche build-up. The tiny current associated with such avalanches can be detected by eliminating the capacitive response of the APD using the circuit shown in Fig.1b, a circuit developed to increase the single-photon count rate.[20] In this circuit the output signal from the APD is split into two paths - one of which introduces a delay of one period of the alternating bias voltage relative to the other path. For the data presented here the APD bias voltage had a frequency of 622 MHz, corresponding to a delay of 1.61 ns. The periodic capacitive response signal from the APD is thus virtually eliminated by taking the difference between the signals in the two paths, as illustrated in Fig.1b. This allows the discrimination of avalanche currents over ten-times weaker than those which can be measured conventionally, as shown in Fig.1c. Meanwhile the photon detection efficiency is similar to that in conventional Geiger mode, increasing up to 30% for an excess voltage of 2.2V.

Figure 2 plots the distribution in the peak output signal voltages from the self-differencing APD for an average detected flux of $\mu$=1.49 photons/pulse from a 1550 nm pulsed laser. Here the excess bias was set to 1.5V, corresponding to a detection efficiency of 10%. The trace shows a series of maxima and shoulders which are centred around 4.7, 8.4, 11.5, 14.6 and 17.6 mV, which we ascribe to the avalanche current induced by 0, 1, 2, 3 or 4 photons. By measuring the avalanche current before it saturates in the device, we can distinguish avalanches stimulated by different numbers of photo-excited holes. Notice that for this particular incident flux (1.49 photons/pulse), the output is dominated by 1- and 2-photon signals, as expected.

To clarify, note that the 0-photon peak in Fig. 2 corresponds to periods for which there is no avalanche, only 'noise' due to an uncompensated component (~4.7 mV) of the capacitive response, due to a small signal mismatch between the two arms of the self-differencing circuit. The dark count rate is negligible (< $2 \times 10^{-6}$ per gate) in all the measurements presented here.

The assignment of the features in Fig. 2 to detection of different photon number states is confirmed by the dependence of the output signal distribution on the incident laser pulse intensity. Figure 3 displays histograms of the peak output signal voltages recorded for different incident intensities. Notice that at the lowest light intensities only the 1-photon peak may be observed along with the 0-photon peak. As the laser intensity increases, the multi-photon peaks become progressively stronger. Notice that, as expected, the probability of 0-photon output decreases with increasing photon flux and the 1-photon output is strongest between 0.80 and 1.54 photons/pulse.

The red lines in Figs. 3a to 3e (and the red line in Fig.2) show a fit to the data assuming a Poissonian distribution of photon number in the incident laser pulses. Gaussians were fitted to the distribution recorded for the lowest incident flux to determine a mean voltage and width for the 0-photon (1-photon) output signal of 4.7 mV (8.4 mV) and 0.175 mV (0.96 mV), respectively. The distribution of the 1-photon output is broader than that of the 0-photon signal due to statistical variations in the size of avalanche triggered by a photon, as discussed in more detail below. The widths of the $N$-photon peaks were then scaled relative to the 1-photon peak according to $N^{0.5}$, while each peak area was fixed according to the expected Poissonian statistics without any fitting parameters.

The data and calculated curves generally show good agreement, except that the mean output voltage of the higher number states are slightly overestimated. This may be due to self-limiting of the stronger avalanches due to the potential drop across the series resistance of the APD. Higher number states may be better resolved by finer balance of the self-differencing circuit, so as to allow the avalanche current and resultant potential drop across the APD to be reduced, or by reducing the series resistance of the APD.

In contrast to the results presented above, we find it is not possible to resolve photon number in conventional gated Geiger mode operation (Fig.1a). For comparison we show in Figs. 3f to 3j the distribution of output signals measured in conventional mode with a gate frequency of 100 kHz and different incident laser pulse intensities. For 0.11 photons/pulse there are two peaks in the distribution corresponding to detection of 0 and 1 or more photons. The light induced peak appears quite asymmetric due to the saturation of the avalanche current for the largest output voltages. Notice that if we increase the flux to 10.6 photons/pulse the distribution does not broaden and no new features appear. On the contrary, the distribution narrows with shrinkage of low amplitude tail, because a larger proportion of the avalanches are saturated at higher photon fluxes. This is in sharp contrast to the behaviour seen in Fig.3a-e, for the self-differencing APD. This confirms that in conventional Geiger mode the APD can only detect the absence or presence of at least one photon, but cannot resolve the number of photons.

It is interesting to compare the self-differencing APD to linear mode operation of APDs at much lower avalanche gain factors (~10), which is often used for low-medium light level detection (without single photon resolution). Unlike in conventional Geiger mode operation, for which the avalanche current is independent of the incident photon number, the self-differencing APD has a linear-mode-like behaviour, ie it produces an avalanche current that is proportional to the illumination flux. Remarkably, however, the multiplication gain of the self-differencing APD is five orders of magnitude higher than is typical in conventional linear mode.

In linear mode the statistical variation in the avalanche gain ($M$) causes error in the measurement of the incident light intensity that is defined by an excess noise factor $F = <M^2>/<M>^2$ and normally increases with $M$ (ref. 21). For InGaAs APDs, the excess noise factor is typically 5.5 at $M = 10$ (ref. 22). Thus even if we could record the very weak output generated by a few photons incident on the device in conventional low-gain linear mode, statistical variations in the gain would prevent photon number resolution.

In Fig. 4 we plot the excess noise ($F$) as a function of mean gain, $<M>$, recorded at different excess biases. Here $F$ is determined from the variance of the Gaussian fitted to the 1-photon signal measured under low illumination conditions. Remarkably, the self-differencing APD maintains a low excess noise, even for $<M>$ beyond $1\times10^6$, which does not increase with the mean gain. Combining high internal gain and low noise, the self-differencing APD could be a replacement for linear mode APDs where high sensitivity at very low illumination level is needed.

To conclude, avalanche photodiodes can resolve photon number. By using a self-differencing circuit for the readout of the APD we were able to measure very small currents at the early stage of avalanche development. We have shown that in this case the total current through the APD scales proportionally to the number of incident photons. This will be useful not only for applications in quantum optics, but also for photon number counting in low light level detection.

*Correspondence and requests for materials should be addressed to andrew.shields@crl.toshiba.co.uk.

**Figure Captions:**

**Figure 1) Comparison of conventional gated Geiger mode with the new self-differencing mode of APD operation**. **a)** Conventional gated Geiger mode. The detector is biased ($V_{bias}$) with short voltage pulses, with a repetition frequency $f$, to a level ($V_{BR}+V_{EX}$) above the avalanche breakdown voltage $V_{BR}$. Photon pulses are synchronised with the gating bias. As the photon-induced output signal is superposed upon the capacitive response of the APD to the applied bias, only strong avalanches can be detected. **b)** Elimination of the capacitive response of the APD with the self-differencing circuit (green box) used in this work allows the detection of unsaturated avalanches, for which the output voltage scales with the number of detected photons. **c)** Comparison of the avalanche currents measured in conventional gated Geiger (dashed line) and for self differencing (solid) mode for an average detected flux of µ=0.01 photons/pulse. Using the self-differencing APD much weaker avalanches may be detected.

**Figure 2) Distribution of the peak output signal generated by the self-differencing APD**. The APD was illuminated by a 1550 nm pulsed laser with an average detected intensity of µ=1.49 photons/pulse. The lines show a fit to the data described in the text.

**Figure 3) Comparison of output signals recorded for self-differencing and conventional mode.** **3a-e)** Distributions of peak output signal generated by the self-differencing circuit (data points) for different incident laser intensities as marked, together with the calculated distribution (red line). **3f-j)** Distributions of the peak APD output signal measured in conventional gated Geiger mode under different levels of illumination.

**Figure 4) Noise factor analysis.** The dimensionless Excess Noise factor ($F$) in self-differencing mode as a function of the mean gain $<M>$.

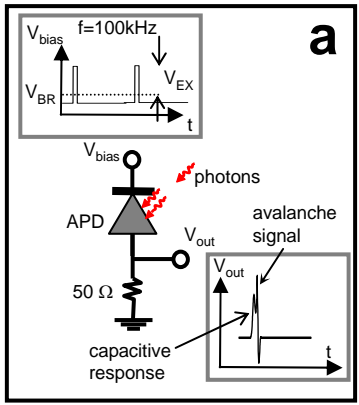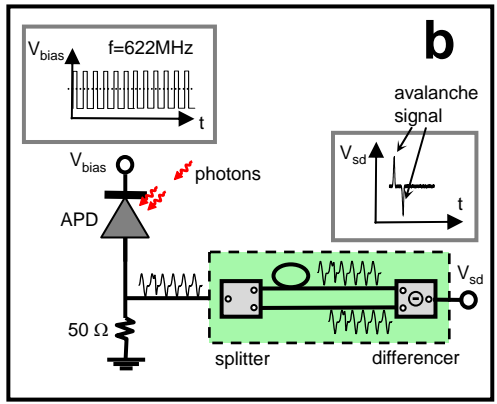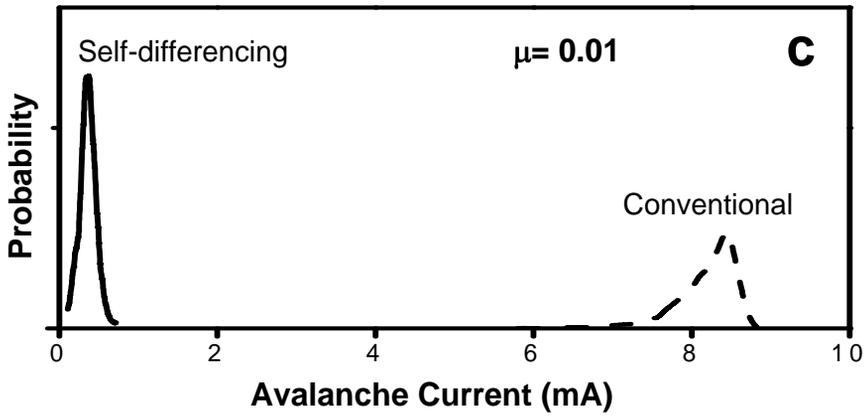

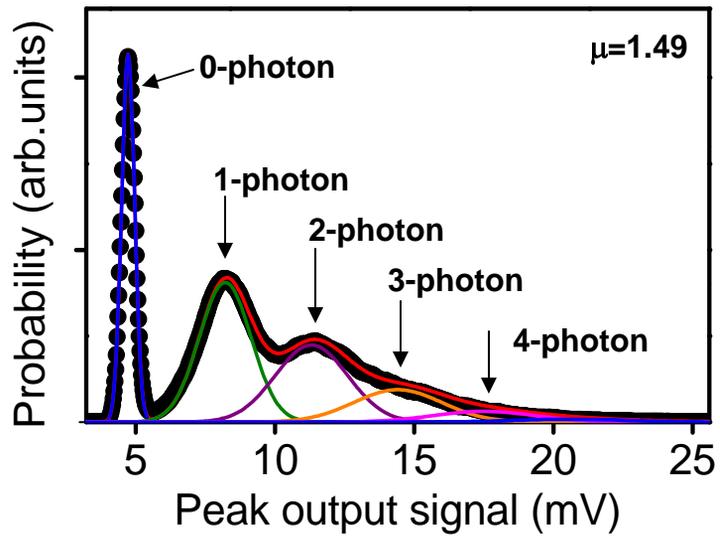

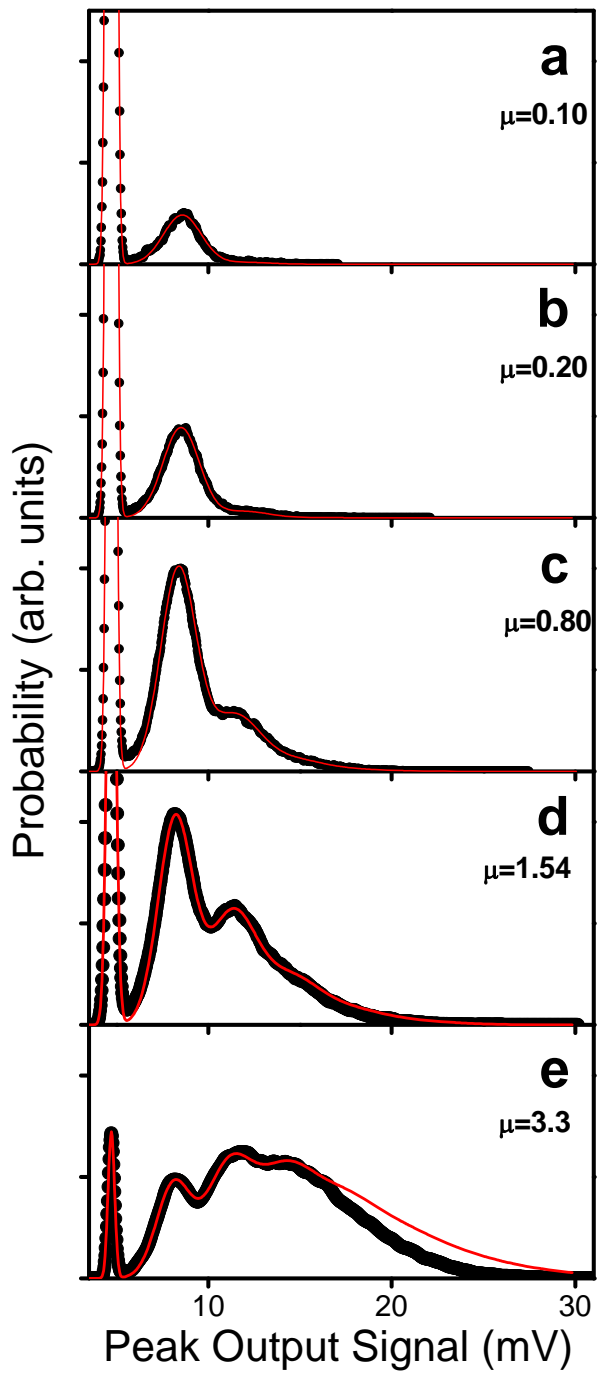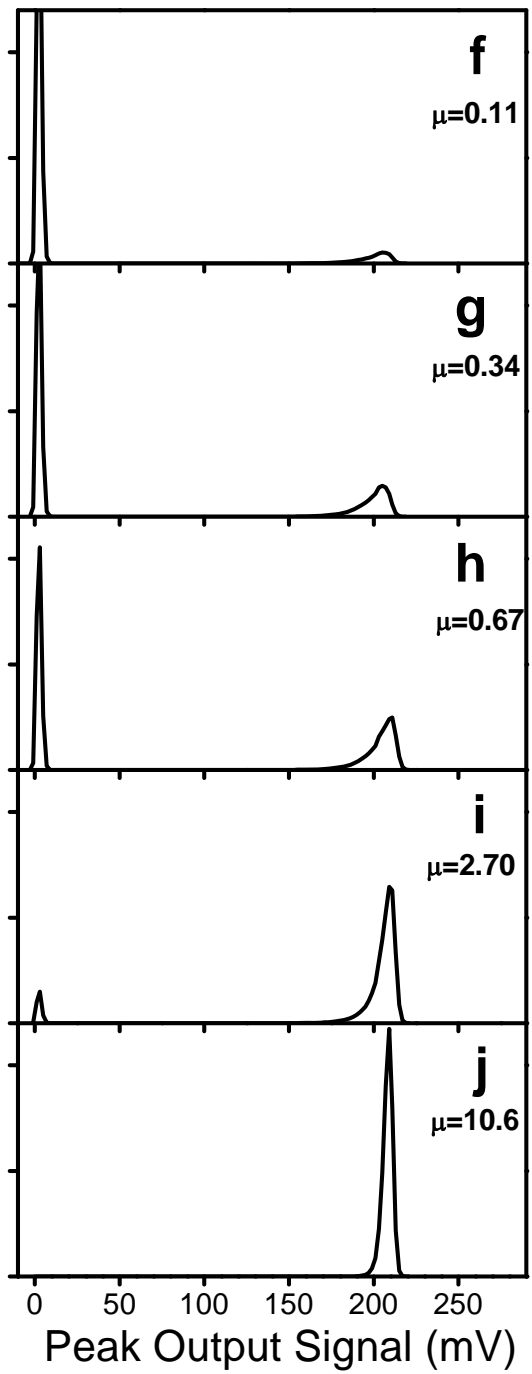

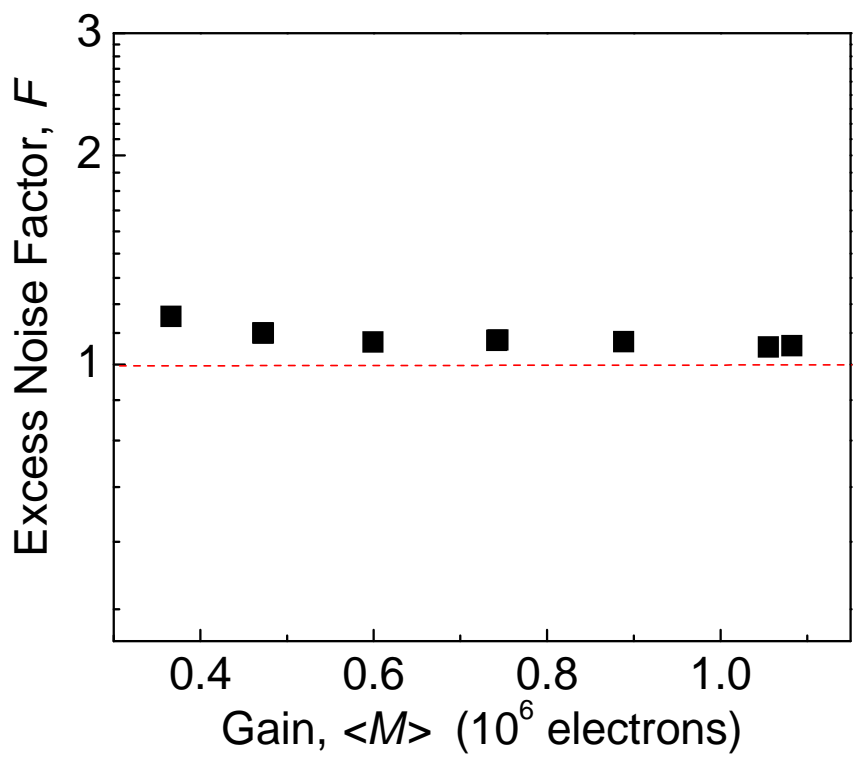